\documentclass{article}
\usepackage[left=2.5cm,right=2.5cm,top=3cm,bottom=3cm,a4paper]{geometry}
\usepackage[numbers]{natbib}
\usepackage{hyperref}
\usepackage{subfiles}
\usepackage{longtable}
\usepackage{graphicx}
\usepackage{multirow}
\usepackage{amsmath,amssymb}

\newenvironment{myindentpar}[1]%
{\begin{list}{}
         {\setlength{\leftmargin}{#1}}
         \item[]
}
{\end{list}}

\begin{document}

\title{Pairwise heuristic sequence alignment algorithm based on deep reinforcement learning}

\author{Yong Joon Song, Dong Jin Ji, Hye In Seo, Gyu Bum Han, and Dong Ho Cho}

\begin{abstract}
Various methods have been developed to analyze the association between organisms and their genomic sequences. Among them, sequence alignment is the most frequently used for comparative analysis of biological genomes. However, the traditional sequence alignment method is considerably complicated in proportion to the sequences' length, and it is significantly challenging to align long sequences such as a human genome. Currently, several multiple sequence alignment algorithms are available that can reduce the complexity and improve the alignment performance of various genomes. However, there have been relatively fewer attempts to improve the alignment performance of the pairwise alignment algorithm. After grasping these problems, we intend to propose a new sequence alignment method using deep reinforcement learning. This research shows the application method of the deep reinforcement learning to the sequence alignment system and the way how the deep reinforcement learning can improve the conventional sequence alignment method.
\end{abstract}

\maketitle

\section{Introduction}
Recent advancements in sequencing technology have enabled the analysis of organisms with long sequences \cite{NGS}. In case of organisms with short sequences, the evolutionary distances between the organisms could be easily analyzed using the older pairwise alignment methods, such as the conventional Needleman--Wunsch (NW) algorithm \cite{NW}, and the relationship between the organisms could be investigated. However, the conventional pairwise alignment method is extremely simple to determine the characteristics of genomic sequences; therefore, it is difficult to balance the complexity and performance. Various attempts have failed in long sequence alignment. The conventional dynamic programming based NW alignment method has a complexity that is proportional to the product of the lengths of the two different nucleotide sequences, which causes great difficulty from the alignment perspective.

The sequence alignment method has reduced the complexity through the development of a multiple sequence alignment (MSA) method that aligns multiple sequences simultaneously. This method has been improved through various approaches, such as detecting common subsequences, ordering the progressive alignment, or increasing the speed using multithreads based on GPUs \cite{MSA1, MSA2, MSA3}. However, there remains a critical point in that the complexity of pairwise alignment could be more severe in case of MSA \cite{CP1}.

In particular, several pairwise alignment algorithms, such as banded alignment, the BLAST, and the MUMmer have been proposed to improve the speed of pairwise alignment \cite{GA2, LA1, GA1}. These alignment methods attempted to solve the complexity issue by limiting the range of the alignment and extending the alignment after word matching or average common substring matching. However, in these cases, there were some accuracy problems in the process of extending short local alignments to large and complex sequences.

To overcome the problems of conventional alignment methods, we proposed a novel alignment method using deep reinforcement learning agent. Reinforcement learning is a way to teach an agent that could choose the best actions by observing an environment in a given system. The conventional tabular based reinforcement learning has a difficulty in expiring and learning the massive and complex systems. To improve this, the deep reinforcement learning method was proposed, and it overcame the limitations by approximately learning the complex systems \cite{DQN1}. The development of reinforcement learning has shown amazing performance in various complex systems \cite{DQN3, DQN2, SSD}. Therefore, we decided to apply this deep reinforcement learning method to sequence alignment system that is willing to find the optimal matches in two complete sequences. Thus, we will describe the application of the deep reinforcement learning to the sequence alignment system in this paper.

\begin{myindentpar}{1em}
The key contribution of this paper is as follows:
\item -Apply the deep reinforcement learning algorithm to the sequence alignment system
\item -Investigate the effect of each parameter on the performance of sequence alignment
\item -Verify alignment performance superiority in short fairly dissimilar sequence pairs
\item -Combine the proposed DQNalign with the conventional sequence alignment algorithm
\item -Prove how the proposed DQNalign can obtain the optimal alignment's performance
\end{myindentpar}

\section{Materials and Methods}

\begin{figure*}
\centerline{\includegraphics[width=14cm]{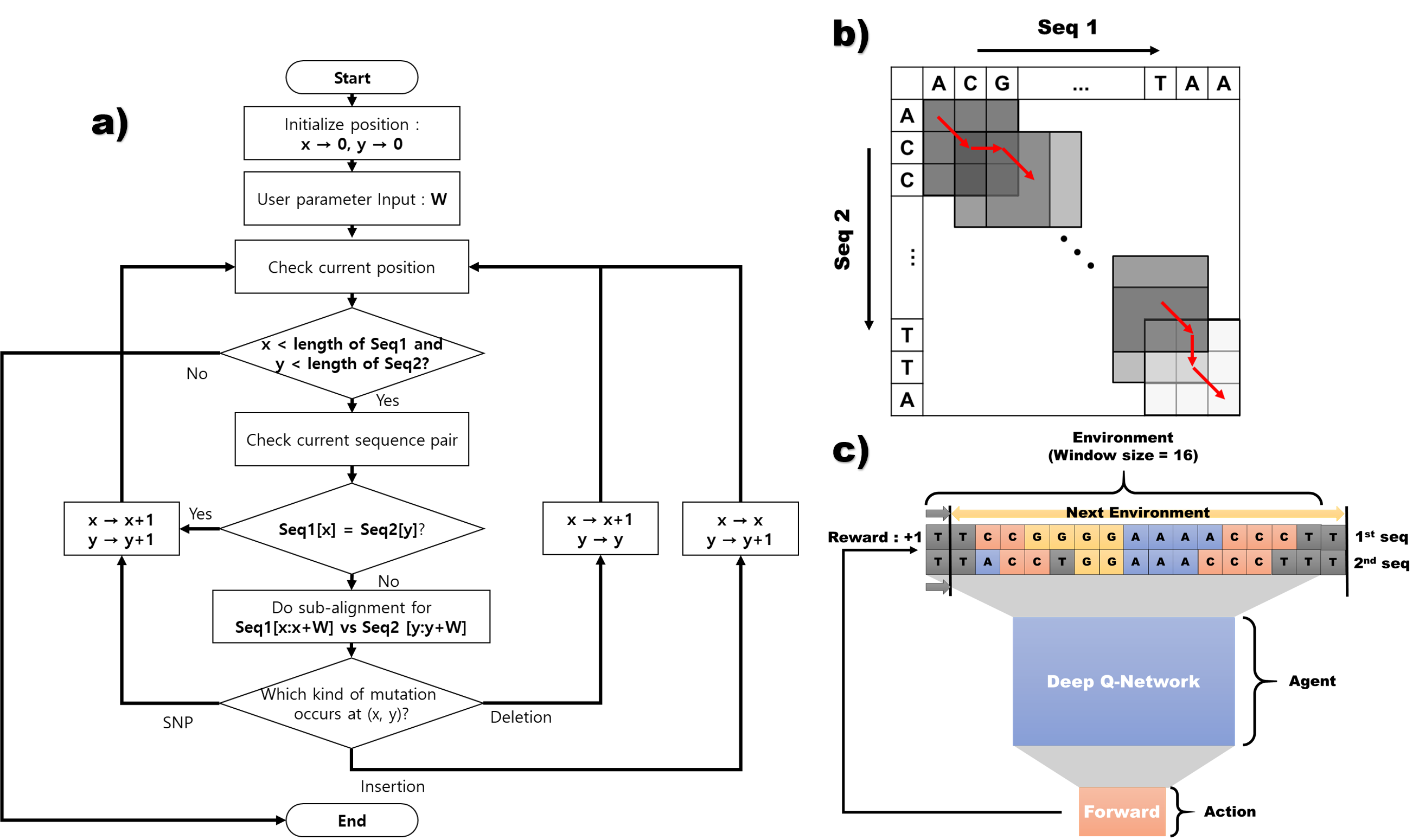}}
\caption[a) Total process of the proposed DQNalign method. DQNalign is divided into two parts. First, b) Local best path selection method, which is a method of repeating window movement by setting direction of progress through alignment process between small sub-sequences. Second, c) deep reinforcement learning-based local best path selection method, which solves high complexity problem of repeating small alignments. Reinforcement learning defines subsequences of current window as environment, and agent is defined as deep reinforcement learning. Here, reinforcement learning proceeds based on rewarding according to scoring strategy of sequence alignment.]{a) Total process of the proposed DQNalign method. DQNalign is divided into two parts. First, b) Local best path selection method, which is a method of repeating window movement by setting direction of progress through alignment process between small sub-sequences. Second, c) deep reinforcement learning-based local best path selection method, which solves high complexity problem of repeating small alignments. Reinforcement learning defines subsequences of current window as environment, and agent is defined as deep reinforcement learning. Here, reinforcement learning proceeds based on rewarding according to scoring strategy of sequence alignment.}
\label{FIG:1}
\end{figure*}

Through this paper, we will introduce the pairwise heuristic sequence alignment algorithm based on deep reinforcement learning. We call this proposed method as DQNalign. The entire procedure of DQNalign is described as the flow chart in Fig.\ref{FIG:1}a. The detailed explanation and the code implementation of DQNalign algorithm are available at \href{https://github.com/syjqkrtk/DQNalign}{https://git hub.com/syjqkrtk/DQNalign}. Here, the proposed DQNalign algorithm can include one deep neural network structure among the two types that are used to predict the processing direction of alignment efficiently : Dueling Double Deep Q-network (DDDQN), faster Dueling Double Deep Q-network (faster DDDQN). The detailed network architecture will be explained in this section.

\subsection{Deep reinforcement learning in sequence alignment system}
To apply deep reinforcement learning to sequence alignment, we try to develop a novel sequence alignment method using reinforcement learning. Instead of observing the entire sequence at once, we propose a novel heuristic sequence alignment method that repeats the small alignment while moving the window of sub-sequence pairs. Through DQNalign, it is possible to solve memory and time complexity problems.

The proposed heuristic sequence alignment method can be expressed as shown in Fig.\ref{FIG:1}b, and we call this proposed heuristic alignment method as local best path selection method. The problem of determining the optimal direction in sub-sequence within a window can also be seen as a kind of sub-alignment process. It can also be inferred that if the window size is expanded to the entire sequence length, the local best path selection method turns into the optimal sequence alignment. We prove this relation between window size and performance through a numerical analysis, which will be discussed in the result section. Also, we propose a novel deep reinforcement learning based sequence alignment system shown in Fig.\ref{FIG:1}c. After 2 sub-sequences in the window are set as environment, the agent with the deep Q-network observes the current environment and selects the direction (forward, insertion, deletion) as the next action. The scoring system of conventional alignment method will be used as reward in reinforcement learning. Through this process, we can execute deep reinforcement learning-based agent that can find the optimal path of the alignment at a given position.

\subsection{Detailed network architecture}
The detail of the deep Q-network based agent is shown in Fig.\ref{FIG:S1} and Fig.\ref{FIG:S2}. Various techniques of the deep reinforcement learning are used for stability and performance. The detailed techniques will be dealt in this section.

\subsubsection{Dueling Double Deep Q-network (DDDQN)}
First, we apply Dueling Deep Q-network \cite{DQN3} and Double Deep Q-network \cite{DQN2} methods to improve the convergence and stability of Deep Q-network (DQN). Dueling DQN is a method that divides the predicted reward (Q value) into a kind of average and variance, and each of them is called as ``Value'' and ``Advantage'', respectively. Using this method, the agent can learn the scores of the states and actions separately, which helps the convergence of the learning progress.

Moreover, Double Deep Q-network method uses a duplicated network, which is called as target network. This target network is used for updating the main network while evading the overestimation. To solve the overestimation problem, Double DQN method uses the target network that converges slowly with preventing the policy from falling into the local minimum. In detail, we make the target network slowly converge towards the main network by constant ratio, tau.

Using these techniques, we define a Dueling Double Deep Q-network as shown in Fig.\ref{FIG:S1}a. Variable window sizes are used as parameter of the convolutional neural networks. Each of the type of nucleotide (A,C,G,T) was converted to a 3 x 3 pixel square with CMYK color. To separate the left, right, top, and bottom end of the sub-sequences, 3 x 3 pixels of empty space are added. The detailed parameters of the network are shown in Fig.\ref{FIG:S1}b. We can see that the number of parameters and FLOPS were linearly proportional to the window size as shown in Table \ref{TAB:1}. Additionally, we also use the experience replay methods to prevent the overestimation.

\begin{table}
\caption[Parameter counts of various network architecture in DQNalign]{Parameter counts of various network architecture in DQNalign}
\begin{center}
\begin{tabular} {c|c|c|c|c}
\hline\hline
\multirow{2}{*}{Window size} &\multicolumn{2}{c|}{DDDQN} &\multicolumn{2}{c}{faster DDDQN}\\
\cline{2-5}
&Param. & FLOPS &Param. & FLOPS \\
\hline
10 &763k& 2.67M&103k&360k\\
30 & 1.68M& 5.88M& 74.0k&259k\\
50 & 2.60M& 9.09M&107k&374k\\
100 & 4.70M& 16.4M&172k&603k\\
\hline\hline
\end{tabular}
\end{center}
\label{TAB:1}
\end{table}

\subsubsection{Faster Dueling Double Deep Q-network: separable convolutional layer based acceleration (faster DDDQN)}
The DQNalign method is not limited to specific network architecture. We try to propose a second version of the network structure focused on reducing complexity. To reduce the complexity, the convolutional layer of the DDDQN is replaced with a separable convolutional layer. We call this modified version of the DDDQN as faster DDDQN. Separable convolutional layer separates a convolutional layer into two different layers called point-wise layer and depth-wise layer \cite{SSD}. This method reduces the number of calculations required for the entire convolutional layer to $1/9$ times in case of 3 x 3 filters.

The modified version of DDDQN, faster DDDQN is shown in Fig.\ref{FIG:S2}a. Like DDDQN, ACGT are mapped into CMYK, and an empty space is applied to the edge. However, in case of faster DDDQN, each nucleotide is mapped to a smaller size of 2 x 2 pixels. Then, we increase the number of layers to 4, unlike the DDDQN. The size of all filters is 3 x 3, and the size of stride is 3, 1, 1, and 3. Then, two maxpooling layers are added to reduce the size of the layer. Detailed parameters are noted in Fig.\ref{FIG:S2}b.

As shown in Table \ref{TAB:1}, we can reduce the number of the operations from $1/9$ to $1/26$ times compared to the DDDQN using separable convolutional layer despite the increased numbers of the layers. Based on these results, we have confirmed that we can make adjustments between complexity and accuracy by controlling the architecture of the neural networks.

We can find the number of the parameters and FLOPS that have an abnormal tendency when the window size is 10. This behavior occurs owing to insufficient size of faster DDDQN, which makes it impossible to add the last convolutional layer. In this case, the faster DDDQN has only three convolutional layers instead of four. Absence of the last convolutional layer causes a large amount of increase in parameters and FLOPS in the fully connected layer. Thus, we can see that the numbers of parameters and FLOPS are large when the window size is 10 in Table \ref{TAB:1}.

\subsection{Training procedure}
Most machine learning requires a massive number of input data. In addition, using only a limited set of data in particular sequences can cause bias in deep reinforcement learning. Therefore, we created new training environments according to the model of evolution.

In the training procedure, we generated the sequence pairs according to following rules. First, we generated a completely random sequence. Then, we made the other sequence by mutating the random sequence. For convenience, we used the JC69 model to create the SNP mutations \cite{NA3}. We used the Zipfian distribution based indel length model for generating the indels \cite{NA4}.

\subsection{Conventional sequence alignment algorithms}
In case of the DQNalign method, the effect of the starting point on the result is critical because the alignment procedure is performed only toward one direction. Therefore, we removed the instability by applying the preprocessing methods of the conventional alignment methods such as longest common substring, the Clustal Omega and the MUMmer. Thus, we will briefly introduce the conventional sequence alignment methods in this section.

\subsubsection{Longest common substring}
Among the preprocessing methods, the longest common substring method is applied first. We extracted the longest common substring using SequenceMatcher function in difflib of python. Using this function, the most similar portion in the sequence pair was found. Then, this longest common substring was used as a starting point of the DQNalign method.

\subsubsection{Clustal Omega}
We try to improve the DQNalign method by adopting the pairwise alignment in the Clustal omega. We referred to the code implementation of the Clustal omega at \href{https://github.com/etetoolkit/ext\_apps/blob/master/src/clustal-omega-1.2.1}{https:// github .com/etetoolkit/ext\_apps/blob/master/src/clustal-omega-1.2.1}. This Clustal Omega quickly checks the k-tuple matches of the two sequences, and records the diagonal position of each match. Thereafter, the algorithm calculates the score of each diagonal, and the diagonals of the top scores are selected to be anchors. However, the connection between the diagonals may be too wide to link them when aligning large sequences. In this study, we aim to enable higher performance alignment by connecting these empty areas in the alignment process through the DQNalign method.

\subsubsection{MUMmer}
We use the official version of the MUMmer software in \href{https://sourceforge.net/projects/mummer/}{https://sourceforge.net/projects/mummer/}. Because the MUMmer is an extension based alignment method, it is difficult to complete the global alignment in case of dissimilar sequences or long sequences. To improve the coverage of the alignment, we apply our DQNalign method in addition to the MUMmer to complete the entire alignment based on the alignment results of the MUMmer.

\section{Results}
To show the feasibility and performance of the overall proposed DQNalign algorithms, we designed the following three simulations: 1. Numerical analysis on the step error probability, 2. Performance difference among DQNalign methods according to parameter changes, and 3. Performance comparison with the conventional alignment methods. Based on these simulations, we will show how the DQNalign method can improve the heuristic sequence alignment performance and how we can adapt the DQNalign method into the conventional alignment methods.

\subsection{Numerical results}
Before applying the deep reinforcement learning, it is necessary to prove clearly whether the heuristic sequence alignment method in Fig.\ref{FIG:1} can perform properly. Therefore, we refer the paper \cite{NA1,NA2} to analyze the step error probability of the local best path selection method. In these papers, the distribution of alignment scores is given by the Gumbel distribution with parameter related to the multiplication of the two sequence lengths.

\begin{equation}
    P(S(m,n) \leq s) \sim \exp(-Kmne^{-\lambda s})
    \label{eq1}
\end{equation}

Here, $S(m,n)$ is the distribution of alignment scores between two sequences with lengths $m$ and $n$. $K$ and $\lambda$ are constants determined by the alignment environment. Using this equation, we proved that the local best path selection method can perform an optimal alignment in case of large window sizes.

\subsubsection{Numerical analysis on step error probability}
Prior to the analysis, we make three major assumptions and consider one constraint. First, it is assumed that the NW algorithm can precisely match the mutation information of the actual sequences. Second, two sequence alignments with the same score confirm that both are correct. Third, the alignment at each step in the local best path selection method can be calculated independently. Conversely, for the constraint, we consider the alignment scoring parameters to prevent the indel preferences. We can derive equations on the error probability in case of similar sequences that have positive alignment scores.

\begin{figure*}
\centerline{\includegraphics[width=17cm]{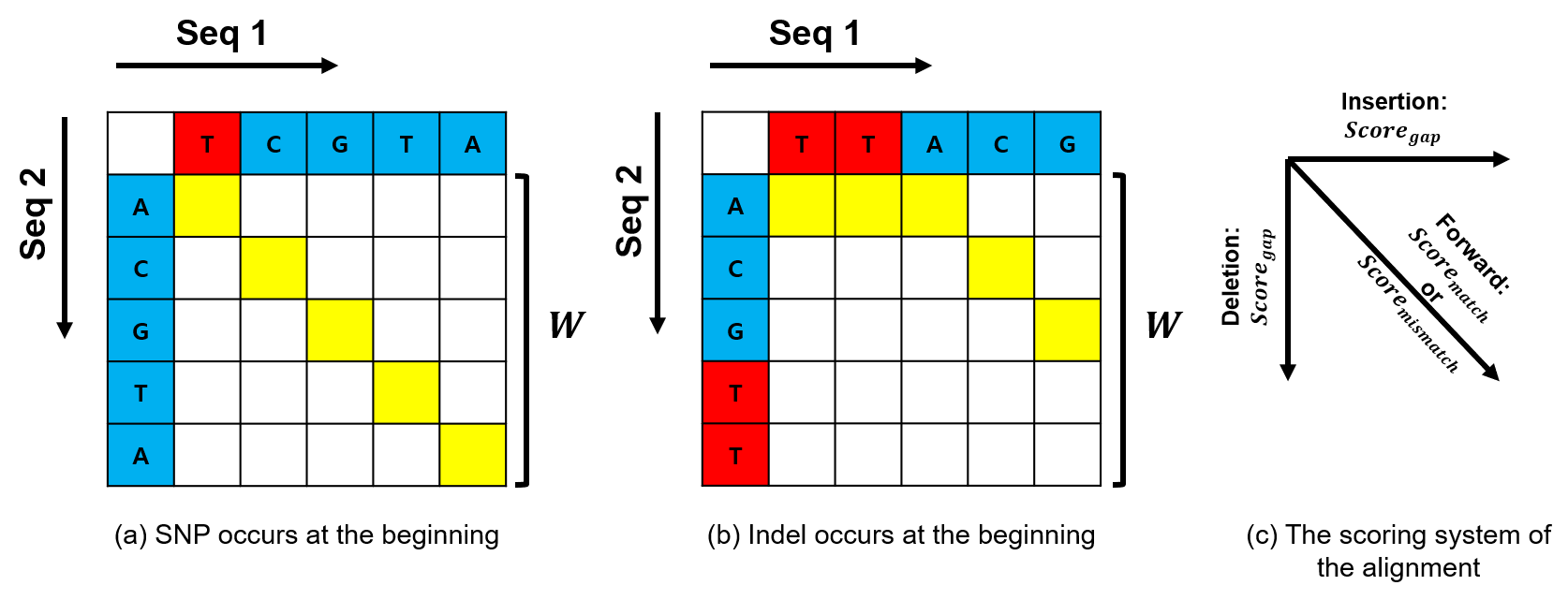}}
\caption[Modeling of local best path selection model for analysis of error probability in case of a fixed window size. a) Expression of SNP within window size. b) Expression of an indel within window size. c) The scoring system of the alignment]{Modeling of local best path selection model for analysis of error probability in case of a fixed window size. a) Expression of SNP within window size. b) Expression of an indel within window size. c) The scoring system of the alignment}\label{FIG:2}
\end{figure*}

Here, we calculate the step error probability of the local best path selection method by considering the SNP and the indel cases seperately. The case of the SNP occurrence is depicted in Fig.\ref{FIG:2}a. If the score of the optimal alignment is $score_{ans}$, we can say that an error occurs when the score of indel direction is higher than $score_{ans}$. In addition, $W$ means the window size, and $score_{gap}$ means the gap penalty of the alignment system. Then, the step error probability can be expressed using the Gumbel distribution as follows.

\begin{equation}
    \begin{aligned}
&P(S(W,W-1) + {score}_{gap} > {score}_{ans})\\
&\simeq 1 - \exp(-KW(W-1)e^{\lambda ({score}_{ans}-{score}_{gap})})
    \end{aligned}
    \label{eq2}
\end{equation}

Here, the $score_{ans}$ can be expressed as $Wscore_{avg}$. Then, the step error probability can be summarized in case of infinitely large W as following. 

\begin{equation}
    \begin{aligned}
        P_{e,SNP} \simeq \lim_{W \to \infty} 2Ke^{\lambda {score}_{gap}} \frac{W^{2}}{e^{\lambda W {score}_{avg}}} \to 0
    \end{aligned}
    \label{eq2-5}
\end{equation}

We confirm that the error probability, $P_{e,SNP}$, converges to 0 when W is infinitely large. The detailed derived process is summarized in supplementary material S1. Based on the process used to derive the step error probability in case that SNP occurs, we can analyze the step error probability, $P_{e,indel}$, for the indel environment in Fig.\ref{FIG:2}b, which is expressed as follows. 

\begin{equation}
    \begin{aligned}
        P_{e,indel} \simeq &K(e^{\lambda {score}_{gap}}+\frac{1}{4}e^{\lambda {score}_{match}}+\frac{3}{4}e^{\lambda {score}_{mismatch}})\\
&\times \lim_{W \to \infty} \frac{W^{2}}{e^{\lambda W {score}_{avg}}} \to 0
    \end{aligned}
    \label{eq4}
\end{equation}

Here, $score_{match}$ means the match score, and $score_{mismatch}$ means mismatch score in the alignment system. Finally, the total step error probability, $P_{total}$, including the occurrence probability of indel, $P_{indel}$ and that of SNP, $P_{SNP}$ is as follows.

\begin{equation}
    \begin{aligned}
        P_{e,total} \simeq &({p}_{indel} K(e^{\lambda {score}_{gap}}+\frac{1}{4}e^{\lambda {score}_{match}} +\frac{3}{4}e^{\lambda {score}_{mismatch}})\\
&+ 2{p}_{SNP}Ke^{\lambda {score}_{gap}}) \frac{W^{2}}{e^{\lambda W {score}_{avg}}} \to 0
    \end{aligned}
    \label{eq2-9}
\end{equation}

\subsubsection{Simulation results to the step error probability analysis}
To verify the step error probability equation through numerical analysis, we designed a simple simulation. In the in-silico simulation, we considered the SNP and indel probability between two sequences in Table \ref{TAB:S2}. Then, we could get the results shown in Fig.\ref{FIG:S3}. Through this simulation, we were able to confirm that the tendencies of simulation and numerical results are similar. In case of large window size, the step error rate converged to near zero.

\subsection{Simulation results among the various types of DQNalign method}
We designed the following three simulations to analyze the performance according to various parameters in the proposed scheme: 1. performance convergence of the deep reinforcement learning-based alignment in the training procedure and 2. performance analysis according to various parameters. Through these simulations, we showed the processes adapting deep reinforcement learning to alignment and considering the optimal parameters of the DQNalign method.

\subsubsection{Performance convergence in the training procedure}
\begin{figure*}
\centerline{\includegraphics[width=17cm]{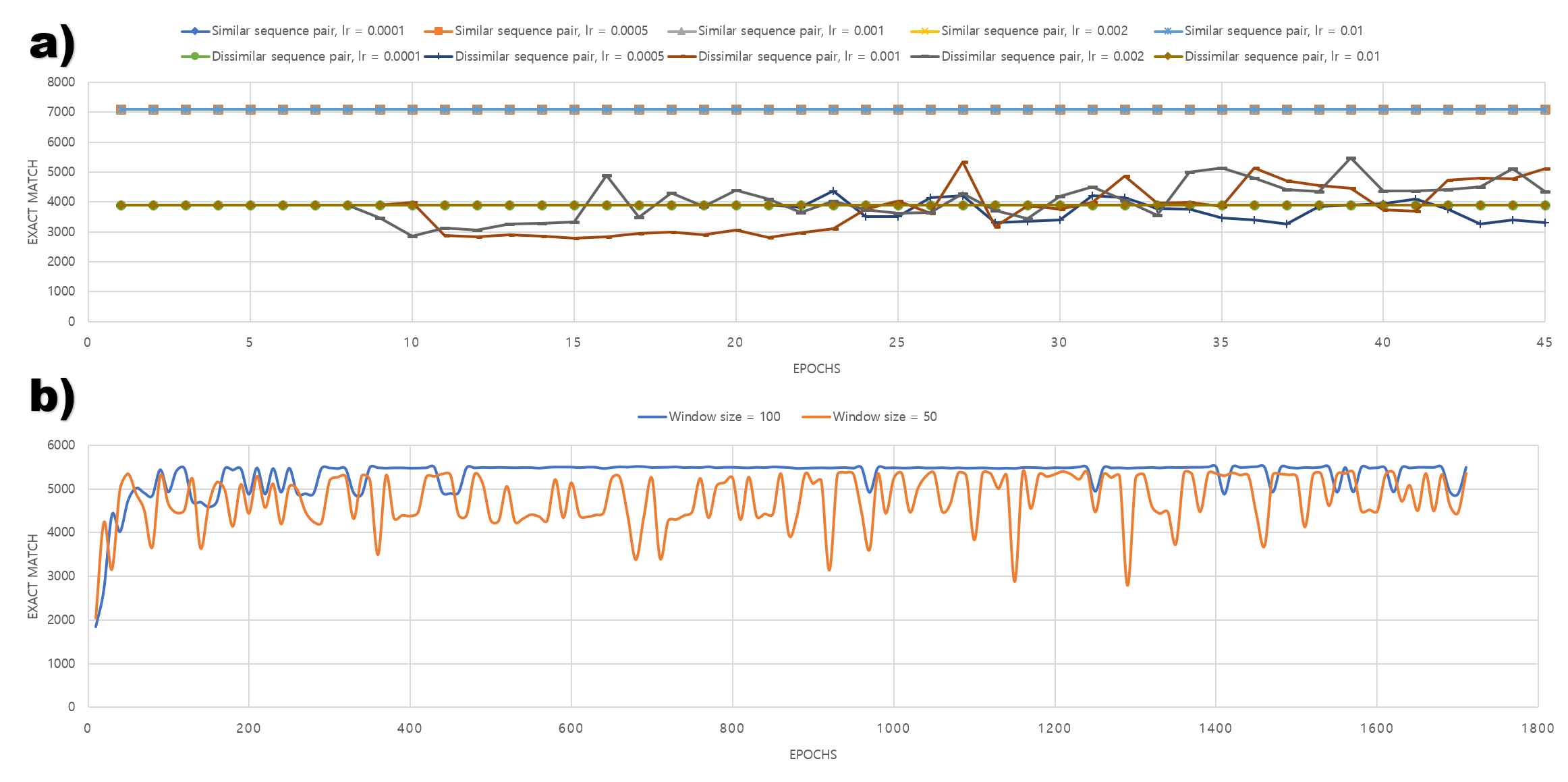}}
\caption[Performance convergence graph of DQNalign method. a) The tendency changes of the convergence according to various learning rates for faster DDDQN in case that the window size is 100. b) The convergence behavior of faster DDDQN in total training procedure in case of dissimilar sequence pair]{Performance convergence graph of DQNalign method. a) The tendency changes of the convergence according to various learning rates for faster DDDQN in case that the window size is 100. b) The convergence behavior of faster DDDQN in total training procedure in case of dissimilar sequence pair}\label{FIG:3}
\end{figure*}

The performance convergence of DQNalign method in training procedure is shown in Fig.\ref{FIG:3}. We prepared two HEV genome sequence pairs to evaluate the training process. One was a similar sequence pair (B1(Bur-82) vs. B2(Bur-86)) and the other was a dissimilar sequence pair (B1(Bur-82) vs. HE-JA1).

In case of similar sequence pair, we could see convergence within a short time regardless of the learning rates in the training procedure. This phenomenon occurs because the environment has rare indel cases in the training procedure. Therefore, the AI learns the wrong behavior that obtains high scores by judging only the forward direction regardless of the environment.

However, it is necessary to learn various environments while evading the biased results to align the actual sequence pairs. Fig.\ref{FIG:3}a shows the test results in a challenging environment in case of dissimilar sequence pairs. As can be seen in the result, the agent could not learn the alignment process in case of improper learning rates, and it can escape the biased results in case of proper learning rates. Therefore, we could figure out that it is crucial to tune the learning rates and the other parameters. Then, we could get the final convergence graph in Fig.\ref{FIG:3}b. Here, by adjusting the learning rates, we could see that the network converge within thousands of epochs. We used these converged networks to evaluate the performance of DQNalign method. Also, you can download the converged network from our Github link.

\subsubsection{Effect of parameters}
\begin{figure*}
\centerline{\includegraphics[width = 17 cm]{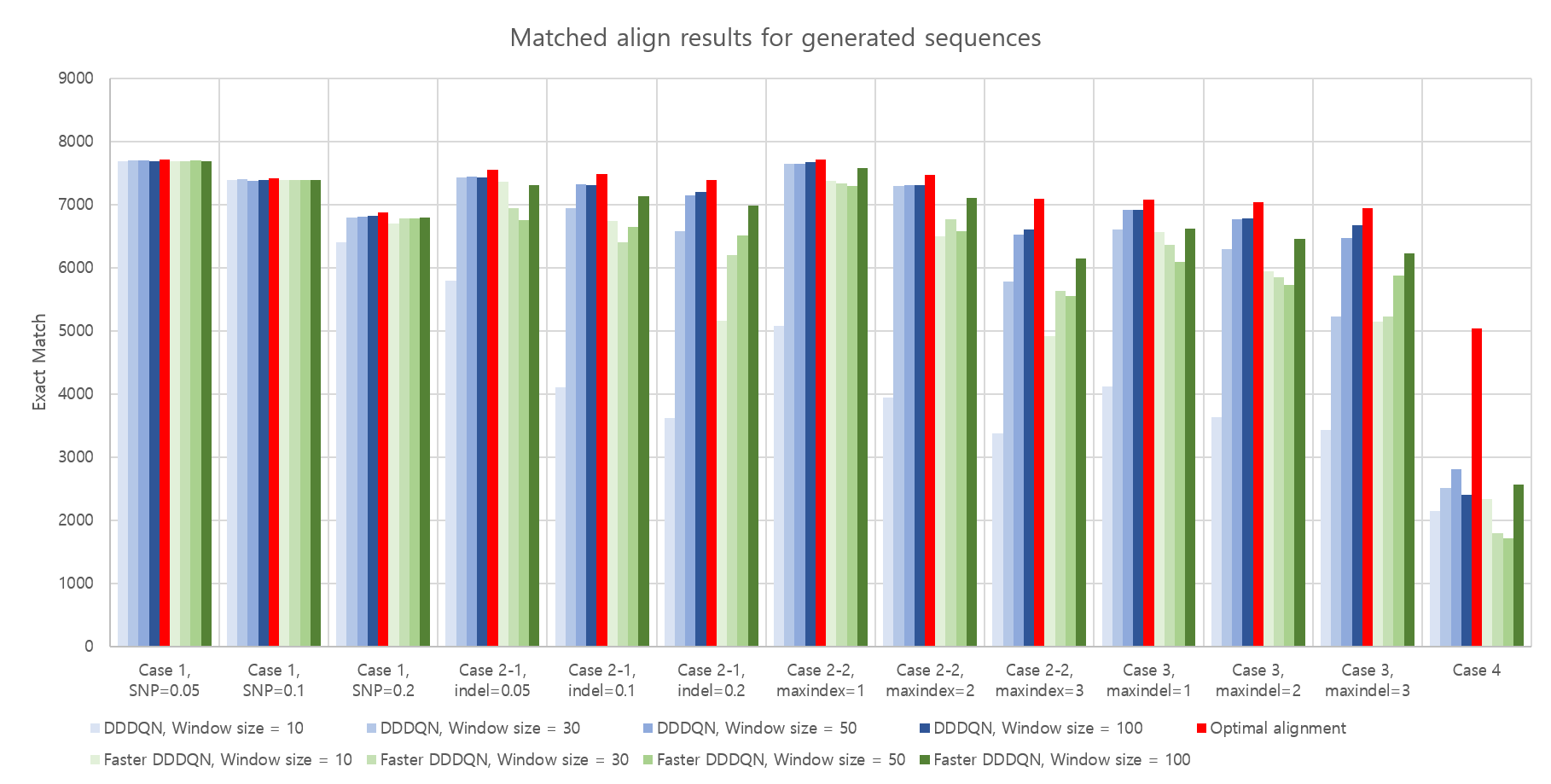}}
\caption[Exact match results of DQNalign method and optimum results in various simulation scenarios]{Exact match results of DQNalign method and optimum results in various simulation scenarios}\label{FIG:4}
\end{figure*}

To observe the effect of the parameters, we used the simulation environments in Table \ref{TAB:S3}. Here, we compared the exact match result of DQNalign method with that of conventional alignment methods. Also, we considered the Needleman--Wunsch algorithm's result to know the ideal outcome of each simulation case. Then, the result of sequence alignment in the actual sequences for the simulation cases in Table \ref{TAB:S3} is shown in Fig.\ref{FIG:4}.

As shown in Fig.\ref{FIG:4}, an increase in the window size increased the alignment performance. When the window size is 100, the performance was converged. Therefore, we could see that 100 is the optimal window size for the HEV sequence set.

Moreover, DQNalign method is not limited to a specific neural network structure, which means that it can learn the alignment process using various types of network structures. As can be seen in this figure, we can confirm that DDDQN has slightly better performance than the faster DDDQN. However, in the complexity results in Table \ref{TAB:2}, it was confirmed that the execution time of faster DDDQN is about two times faster than DDDQN. Based on these results, we were able to check the pros and cons of the proposed alignment networks. We also used the optimal window size up to 100 for each network in the next section to treat the actual sequence data.

\begin{table}
\caption[Average time spent in full alignment of HEV genome cases for various networks (s)]{Average time spent in simulation for various networks (s)}
\begin{center}
\begin{tabular} {c|cccc}
\hline\hline
Window size & 10 & 30 & 50 & 100 \\
\hline
DDDQN& 1.916 & 2.910 & 2.522 & 2.911\\
faster DDDQN& 1.080 & 1.327 & 1.302 & 1.411\\
\hline
Optimal alignment &\multicolumn{4}{c}{143.6}\\
\hline\hline
\end{tabular}
\end{center}
\label{TAB:2}
\end{table}

\subsection{Comparison with conventional alignment algorithms}
To confirm the difference between the DQNalign method and the conventional sequence alignment algorithm, two sequence sets were used: 1. HEV genome sequence set 2. two E.coli genome sequences. In this section, we compared the exact match performance and time complexity between various sequence alignment methods: Clustal omega, MUMmer, and the DQNalign method. The used detailed parameters are described in Table \ref{TAB:S5} and Table \ref{TAB:S6}.

\subsubsection{Simulations on the HEV genome sequence set}
\begin{figure*}
\centerline{\includegraphics[width=17cm]{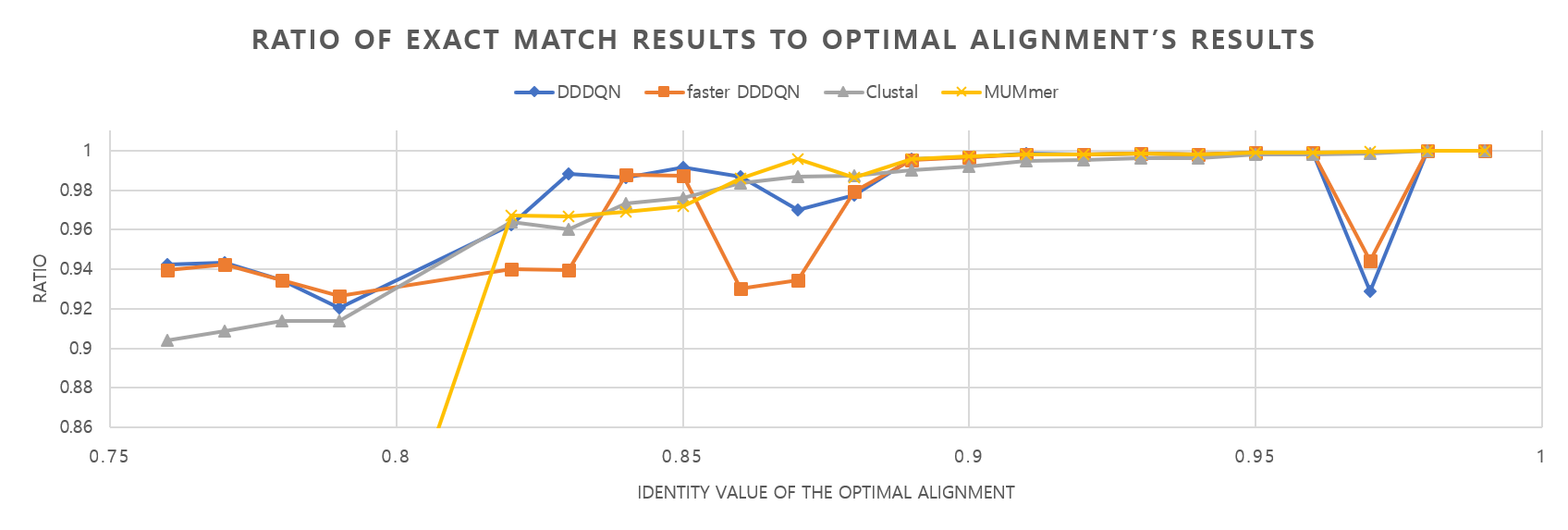}}
\caption[Exact match results in real sequences, which consist of 47 Hepatitis E Viruses. In this sequence set, there are 1081 sequence pairs.]{Exact match results in real sequences, which consist of 47 Hepatitis E Viruses. In this sequence set, there are 1081 sequence pairs.}\label{FIG:5}
\end{figure*}

The results of the HEV sequence set in Table \ref{TAB:S4} are shown in Fig.\ref{FIG:5}. All the alignment results are attached to the supplementary material S3. The ratio of exact matches in the heuristic alignment methods against optimal alignment was used for performance measure. As shown in Fig.\ref{FIG:5}, we could confirm that DQNalign and conventional methods show similar results to the optimal alignment for sequence pairs, which have higher identity value of above 0.88.

However, in the lower identity range, the difference between the alignment methods began to increase. When the identity value of the alignment is from 0.8 to 0.88, it was confirmed that DDDQN and the MUMmer remain stable at a reasonably high level, and faster DDDQN shows the most unstable results, but still shows more than 93\% performance of the optimal alignment. However, when the identity value of the alignment is less than 0.8, which occupies 69.3\% of the entire sequence pairs, DDDQN and faster DDDQN showed noticeably higher performance than the conventional alignment methods. The main reason for this difference is that in the case of dissimilar sequence pairs, the rapid decrease of the number of anchors (Ex. k-tuples in the Clustal method and MUMs in the MUMmer) causes the failure in the process connecting the anchors. Therefore, the conventional alignment methods could not complete the entire alignment and showed low coverage and exact matches. However, DQNalign method immediately estimated the window and decided the direction of alignment regardless of the anchors. Hence, the proposed DQNalign method has less difficulty in aligning the sequence pairs even in case of relatively low identity values.

\subsubsection{Simulations on the E.coli genome sequence set}
To observe the possibility of the alignment in long sequences, we did simulations on Escherichia coliO 157 and Escherichia coli K-12. These alignment results are contained in Supplementary material S3. The brief alignment results are shown in Table \ref{TAB:3}.

\begin{table*}
\caption[Alignment results of the E.coli sequences: Escherichia coli O157:H7 str. TW14359 vs. Escherichia coli str. K-12 substr. MG1655]{Alignment results of the E.coli sequences: Escherichia coliO157:H7 str. TW14359 vs. Escherichia coli str. K-12 substr. MG1655}
\begin{center}
\begin{tabular} {cccccc}
\hline\hline
&DQNalign& Mummer &DQNalign with Mummer &Clustal&DQNalign with Clustal\\
\hline
Exact matches & 2359437 & 3990344 & 4132349 & 3890193  & 4088757 \\
\hline
Consumed time (s) & 35133 & 7.13 & 7.13 + 692 & 514139 & 514139 + 632\\
\hline\hline
\end{tabular}
\end{center}
\label{TAB:3}
\end{table*}

As shown in Table \ref{TAB:3}, the proposed DQNalign has a lower exact match score compared to the other conventional methods owing to the insufficient pre-processing. As expected, sequence alignment for E.coli was not easy in case of using our DQNalign method. There are gene-scale genomic variations in the E.coli samples; therefore, it was hard to connect the whole sequences using only a few hundred-sized windows. Hence, we decided to additionally use the pairwise alignment methods like the Clustal and MUMmer . Here, the overall improved genome alignment result could be obtained by combining conventional alignment methods (the Clustal and MUMmer) with our DQNalign method. We used the DQNalign method to align unaligned gaps of the alignment results obtained using the Clustal or MUMmer methods. Using the DQNalign with Clustal and the DQNalign with MUMmer, we could properly align the complete E.coli sequences.

In view of computational complexity, DQNalign method could align the entire sequence within a few hours, which does not indicate high exact match results but is ten times faster than Clustal method. Also, DQNalign with the Clustal method improved the exact match performance by consuming 0.12\% additional time compared to the conventional Clustal algorithm. However, we confirmed that the MUMmer has a vastly high speed compared to other methods because the MUMmer method focuses on speed optimization in alignment between close sequence pairs.

\section{Discussion and Conclusion}
In this paper, we proposed deep reinforcement learning based alignment method to reduce the complexity of pairwise alignment. Using only simple preprocessing based on the longest common substring (LCS), the DQNalign can align the entire sequences by repeating small alignments. We have confirmed that the DQNalign can achieve high accuracy and low complexity by adjusting the window size adaptively through the analysis.

We confirmed the advantages and disadvantages of the DQNalign method by analyzing the generated sequences and real HEV sequences. We observed that the DQNalign method has more accurate alignment performance than the conventional methods in dissimilar sequence pairs. We confirmed that the MUMmer method has a much faster speed than our DQNalign method. It is desirable to choose the DQNalign or the conventional alignment method in the consideration of real-time processing and accuracy.

The proposed DQNalign method showed the possibility of aligning two long sequences by combining conventional alignment algorithms. It was revealed that the sequences with millions of bases, such as E. coli, could be aligned by combining the DQNalign methods with the Clustal or the MUMmer. Through simulation results, we confirmed that the DQNalign method could be combined efficiently with conventional MUMmer and Clustal methods in view of improving accuracy. 

In the future, we will attempt to apply the DQNalign to local alignment and multiple sequence alignment using improved learning methods and proper training strategies for each case. The DQNalign method used deep reinforcement learning based selection rather than human-based features; therefore, the alignment result was not stable for some sequence pairs. Hence, we will improve the stability of the DQNalign method using the alignment results of the several agents that are differently converged by deep reinforcement learning method. Besides, there were several challenges in securing performance on actual sequences, which result from the gap between the modeled sequences and the real sequences used in the learning. Therefore, we are going to solve these issues using a learning method that directly reflects the actual sequences.



\clearpage
\newcommand{\hbSupplementaryPrefix}{S}
\newcommand{\hbAppendixPrefix}{A}
\renewcommand{\thesection}{\hbSupplementaryPrefix\arabic{section}}
\setcounter{section}{0}
\renewcommand{\thefigure}{\hbAppendixPrefix\arabic{figure}}
\setcounter{figure}{0}
\renewcommand{\thetable}{\hbSupplementaryPrefix\arabic{table}}
\setcounter{table}{0}
\setcounter{equation}{0}
\section{Detailed numerical analysis on local best path selection model}
In this section, we will provide detailed numerical analysis of the error probability of the alignment according to the change of window sizes. To analyze the error probability, problem, three assumptions, and basis of the numerical analysis will be defined. Then, we will prove the convergence of the error probability in case of a large window size.

\subsection{Problem definition}

In this paper, it is necessary to prove that our proposed method approached the optimal value as the window size increased, which is enabled by calculating the error probability at each step. According to previous reports, the probability distribution of the local and global alignment score follows the Gumbel distribution for two sequences with length n and m \cite{NA1,NA2}. Therefore, we derived the equation that corresponds to the distribution of the alignment score and demonstrated how the increase of window size can improve the accuracy of the entire sequence alignment. As we mentioned in the main manuscript, we made three major assumptions and one constraint. First, it is assumed that the NW algorithm can precisely match the mutation information of actual sequences. Second, it is assumed that two sequence alignments with the same score indicate that both are correct. Third, the alignment at each step in the proposed algorithm can be calculated independently. Conversely, for the constraint, we consider the alignment scoring parameters to prevent the indel preferences. Through these assumptions and constraint, we can derive the error probability in case of similar sequences that have positive alignment scores.

\begin{figure}
\centerline{\includegraphics[width = 11 cm]{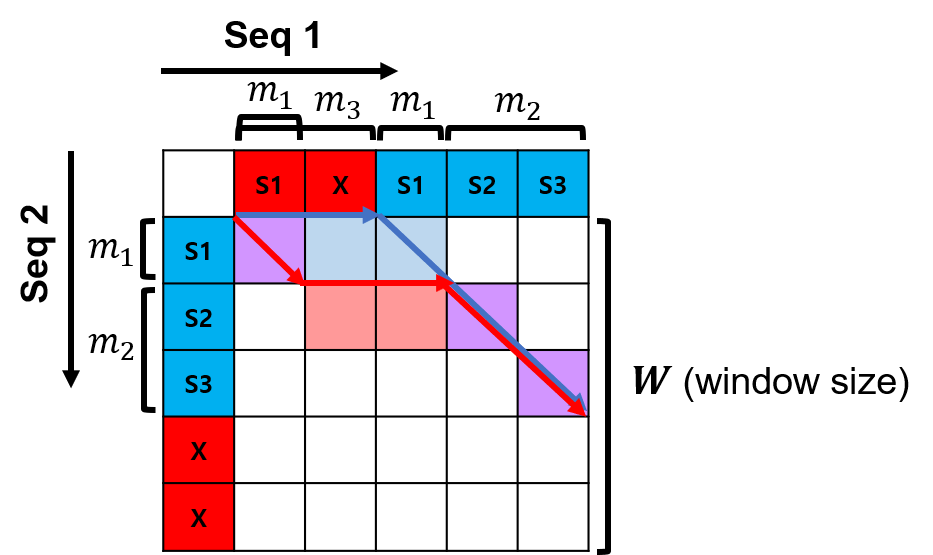}}
\caption[Conceptual diagram of local best path selection model for analysis of error probability in case of fixed window size]{Conceptual diagram of local best path selection model for analysis of error probability in case of fixed window size}\label{FIG:A1}
\end{figure}

We want to explain why the DQNalign skips sub-alignment when the two sequences have the same subsequences in the start of the window, as shown in Fig.\ref{FIG:1}a. Let us consider the case where indel occurred in sequence 1, as shown in Fig.\ref{FIG:A1}. In this case, this indel sequence is identical to $m_1$ of the original sequence at the beginning. Here, the ``red $S1$'' and ``blue $S1$'' have the same nucleotides with length $m_1$. Next, $m_2$ denotes the length of the rest of the identical subsequences (blue subsequences), $m_3$ means the length of the indel (red subsequences), and the notation, $X$ denotes ``don't care'' region.

Let us focus on the two paths shown in this figure. The blue path denotes the ground-truth alignment of the sequence pair, and the red path denotes the result of the sequence alignment obtained by our local best path selection method. Two sequence alignments have the same score, ${score}_{match} \times ({m}_{1}+m_{2}) + {score}_{gap} \times {m}_{3}$. Based on the second assumption, we can regard the red path as the correct path. Also, if we separate these two paths into two segments, the first segment can be seen as a small sequence alignment in the block with ${m}_{1}\times({m}_{1}+{m}_{3})$ subsequences, and the shared remaining paths, ${m}_{2}\times{m}_{2}$, are treated as the second segment.

We could demonstrate that these two paths have the same alignment score even if we consider only the first segment. In the first segment, the best score can be represented as ${m}_{1}{score}_{match}+{m}_{3}{score}_{gap}$. This value is equal to the red path score; therefore, we could state that the red path is also the best path. Therefore, we can infer that the forward alignment offers the best solution when the two sequences are equal at the beginning of the sequences regardless of the indel state. Hence, the proposed DQNalign skips the alignment process when the two sequences have the same subsequences in the start of the window. The detailed algorithm is depicted in Fig.\ref{FIG:1}a.

\subsection{Numerical analysis}

When an error occurs in the local best path selection method with a fixed window size, we can narrow down the number of cases into two cases, as shown in Fig.\ref{FIG:2}(a) and (b). In Fig.\ref{FIG:2}(a), where SNP occurs at the first base pair in the window, the forward direction must be selected. However, when the local best path selection method takes a wrong insertion or deletion direction, it is assumed that an error occurs. In this case, let us consider the Gumbel distribution. All the notations of the equations are mentioned in the main article. The error probability that the score of the insertion or deletion path is larger than the actual score can be obtained as follows.

\begin{equation}
    P(S(W,W-1) + {score}_{gap} > {score}_{answer}) \simeq 1 - \exp(-KW(W-1)e^{\lambda ({score}_{answer}-{score}_{gap})})
    \label{eq2-2}
\end{equation}

\noindent Also, the total error probability $P_{e,SNP}$ that any one of the two cases will occur can be calculated as follows.

\begin{equation}
    P_{e,SNP} \simeq 1 - \exp(-KW(W-1)e^{\lambda ({score}_{answer}-{score}_{gap})})^{2} = 1 - \exp(-2KW(W-1)e^{\lambda ({score}_{answer}-{score}_{gap})})
    \label{eq2-3}
\end{equation}

\noindent For a very large W, the above equation can be written as follows.

\begin{equation}
    \begin{aligned}
        P_{e,SNP} \simeq \lim_{W \to \infty} 1 - \exp(-2KW(W-1)e^{\lambda ({score}_{answer}-{score}_{gap})}) \simeq 1 - \exp(-2Ke^{\lambda {score}_{gap}} \lim_{W \to \infty} \frac{W^{2}}{e^{\lambda W {score}_{avg}}})
    \end{aligned}
    \label{eq2-4}
\end{equation}

\noindent When the average score per nucleotide is ${score}_{avg}$ and if it is greater than 0, we can see that $\lim_{W \to \infty} \frac {{W}^{2}}{{e}^{\lambda W{score}_{avg}}} \to 0$. Then, Eq.\ref{eq2-4} can be described approximately as follows.

\begin{equation}
    \begin{aligned}
        P_{e,SNP} \simeq \lim_{W \to \infty} 1 - \exp(-2Ke^{\lambda {score}_{gap}} \frac{W^{2}}{e^{\lambda W {score}_{avg}}}) &\simeq \lim_{W \to \infty} 1 - (1 - 2Ke^{\lambda {score}_{gap}} \frac{W^{2}}{e^{\lambda W {score}_{avg}}}) \\
&= \lim_{W \to \infty} 2Ke^{\lambda {score}_{gap}} \frac{W^{2}}{e^{\lambda W {score}_{avg}}} \to 0
    \end{aligned}
    \label{eq2-5}
\end{equation}

Using a similar method used in the derivation of Eq.\ref{eq2-2} to Eq.\ref{eq2-5}, we can derive the error equations for the indel case. For the indel case, all of the cases that contain match, mismatch, and the other side indel are treated as errors. Then, we can derive the probability of each case expressed by the Gumbel distribution as follows.

\begin{equation}
    \begin{aligned}
        P(S(W,W-1) + {score}_{gap} > {score}_{answer}) & \simeq 1-\exp(-KW(W-1)e^{-\lambda ({score}_{answer}-{score}_{gap})}) \; &\textrm{in other side indel case} \\
& \simeq 1-\exp(-K(W-1)^{2}e^{-\lambda ({score}_{answer}-{score}_{match})}) \; &\textrm{in match case}\\
& \simeq 1-\exp(-K(W-1)^{2}e^{-\lambda ({score}_{answer}-{score}_{mismatch})}) \; &\textrm{in mismatch case}\\
    \end{aligned}
    \label{eq2-6}
\end{equation}

\noindent Here, we assume the probabilities of a match and mismatch to be 1:3, then the total error rate $P_{e,indel}$ is expressed as follows.

\begin{equation}
    \begin{aligned}
        P_{e,indel} \simeq 1&-\exp(-KW(W-1)e^{-\lambda ({score}_{answer}-{score}_{gap})}) \\
& \times (\frac{1}{4}\exp(-K(W-1)^{2}e^{-\lambda ({score}_{answer}-{score}_{match})}) + \frac{3}{4}\exp(-K(W-1)^{2}e^{-\lambda ({score}_{answer}-{score}_{mismatch})}))
    \end{aligned}
    \label{eq2-7}
\end{equation}

\noindent For a very large W, the above equation can also be summarized as follows.

\begin{equation}
    P_{e,indel} \simeq K(e^{\lambda {score}_{gap}}+\frac{1}{4}e^{\lambda {score}_{match}}+\frac{3}{4}e^{\lambda {score}_{mismatch}}) \times \lim_{W \to \infty} \frac{W^{2}}{e^{\lambda W {score}_{avg}}} \to 0
    \label{eq2-8}
\end{equation}

\noindent When the rates of the SNP and indel are ${p}_{SNP}$ and ${p}_{indel}$, respectively, the total error probability, $P_{e,total}$ can be expressed as follows.

\begin{equation}
    \begin{aligned}
        P_{e,total} \simeq ({p}_{indel} K(e^{\lambda {score}_{gap}}+\frac{1}{4}e^{\lambda {score}_{match}}+\frac{3}{4}e^{\lambda {score}_{mismatch}}) + 2{p}_{SNP}Ke^{\lambda {score}_{gap}}) \times \lim_{W \to \infty} \frac{W^{2}}{e^{\lambda W {score}_{avg}}} \to 0
    \end{aligned}
    \label{eq2-9}
\end{equation}

\clearpage
\renewcommand{\thefigure}{\hbSupplementaryPrefix\arabic{figure}}
\setcounter{figure}{0}

\begin{figure*}
\centerline{\includegraphics[width=17cm]{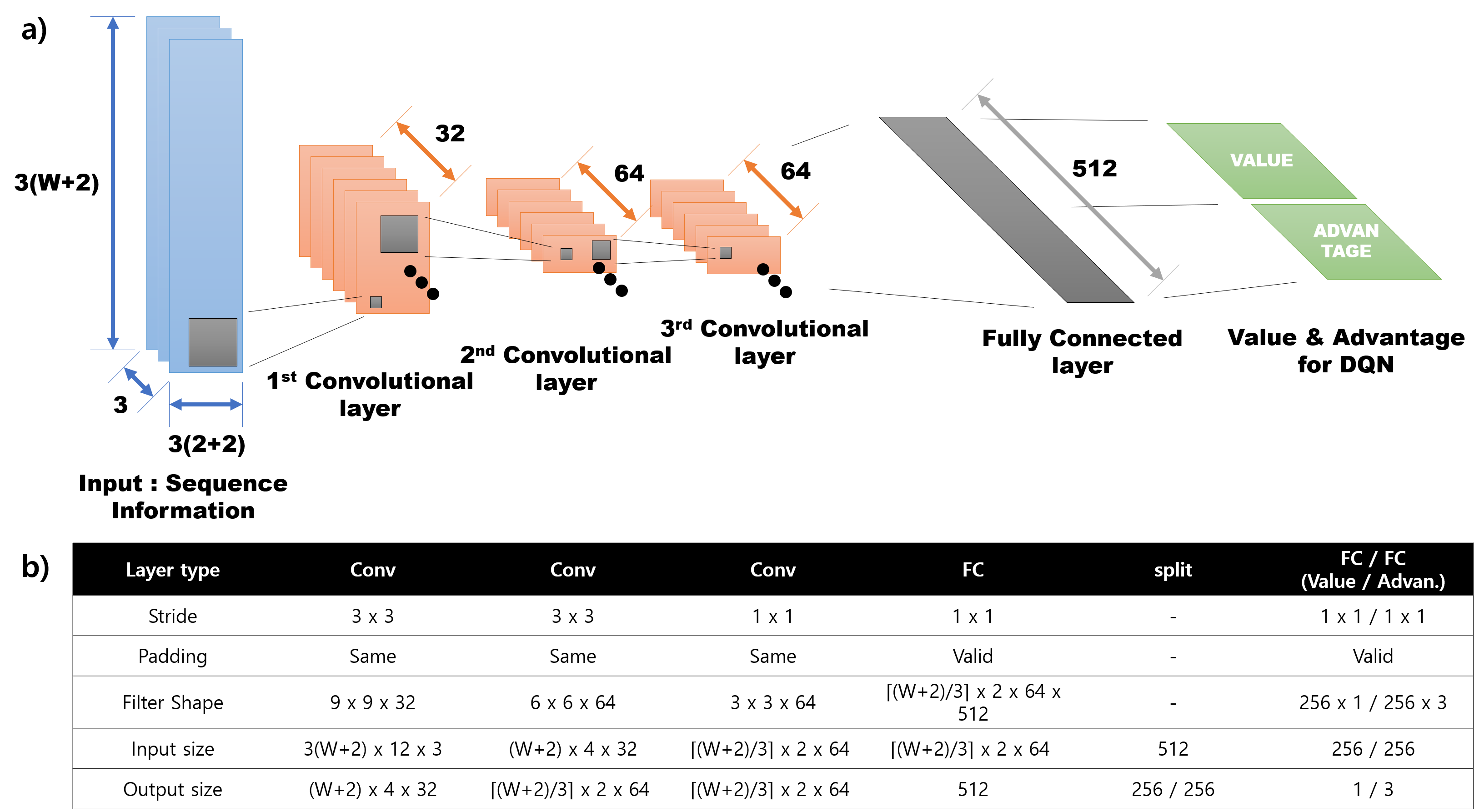}}
\caption[Detailed network architecture of Dueling Double Deep Q network (DDDQN). a) Conceptual diagram of network. b) Parameter table of network]{Detailed network architecture of Dueling Double Deep Q network (DDDQN). a) Conceptual diagram of network. b) Parameter table of network}
\label{FIG:S1}
\end{figure*}

\begin{figure*}
\centerline{\includegraphics[width = 17 cm]{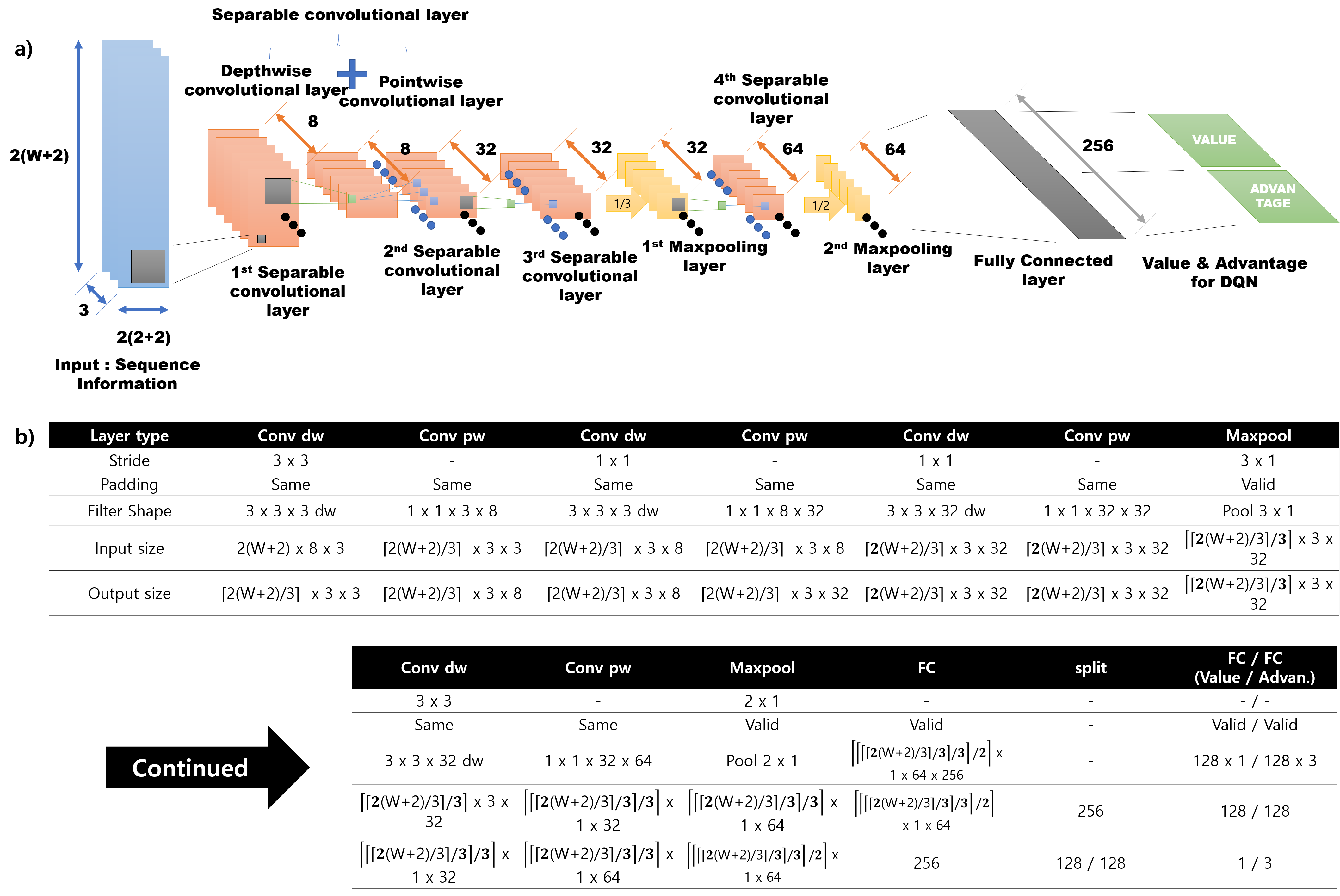}}
\caption[Detailed network architecture of separable convolutional layer based network (faster DDDQN). a) Conceptual diagram of network. b) Parameter table of network]{Detailed network architecture of separable convolutional layer based network (faster DDDQN). a) Conceptual diagram of network. b) Parameter table of network}
\label{FIG:S2}
\end{figure*}

\begin{figure*}
\centerline{\includegraphics[width = 12 cm]{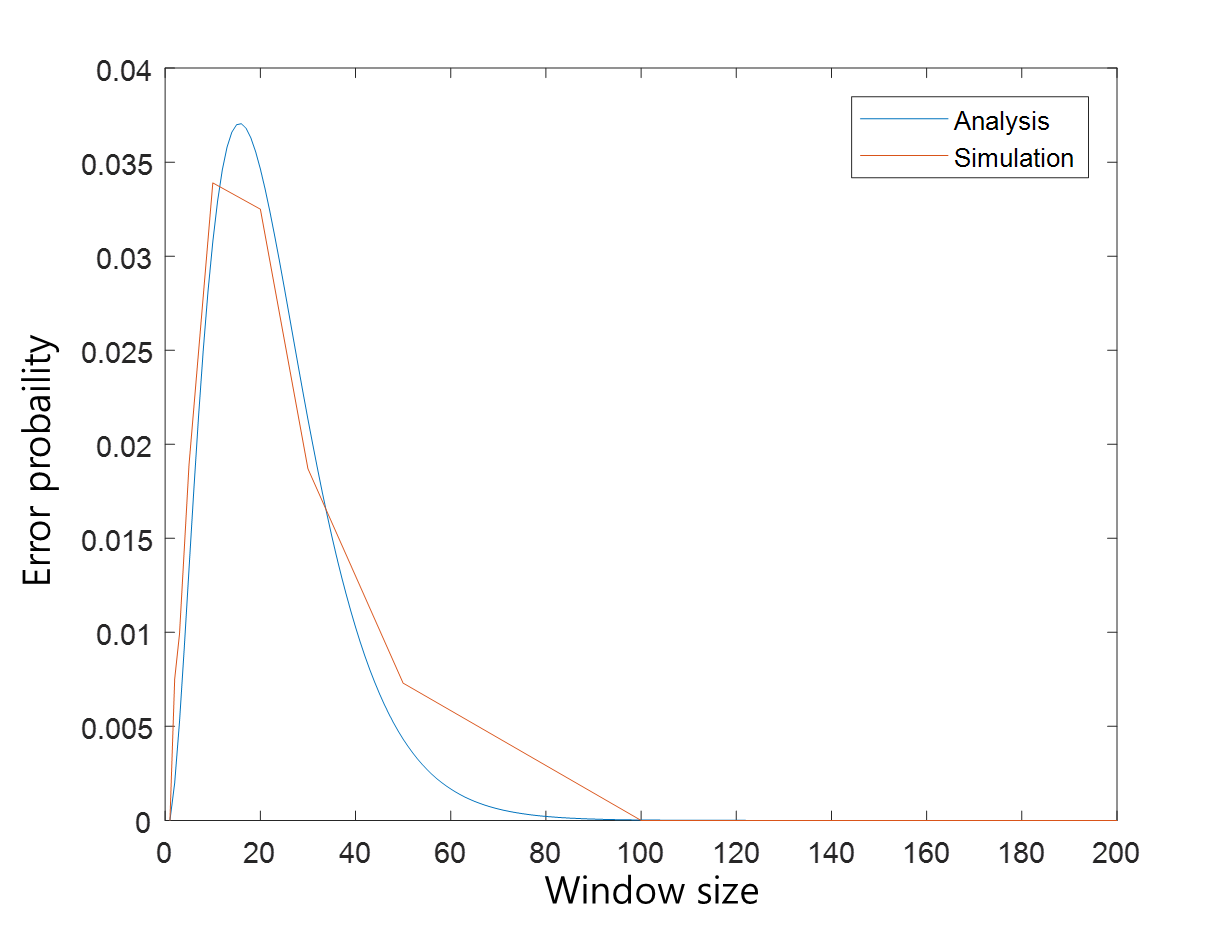}}
\caption[Step error probability according to windows size]{Step error probability according to windows size}
\label{FIG:S3}
\end{figure*}

\clearpage

\begin{table*}
\caption[Detailed parameters of sequence generation for training procedure]{Detailed parameters of sequence generation for training procedure}
\begin{center}
\begin{tabular} {c|c}    
\hline\hline
Parameters & Training environment \\
\hline
\footnotesize Sequence length ($l$) & 8000 \\
\footnotesize Probability of SNP ($p_{SNP}$) & 0.1 \\
\footnotesize Probability of indel ($p_{indel}$) & 0.02  \\
\footnotesize Maximum length of indel ($I_{max}$) & 10  \\
\footnotesize Zipfian distribution parameter ($s$) & 1.6 \\
\hline\hline
\end{tabular}
\end{center}
\label{TAB:S1}
\end{table*}

\begin{table*}
\caption[Detailed parameters of sequence generation for validation of numerical analysis]{Detailed parameters of sequence generation for validation of numerical analysis}
\begin{center}
\begin{tabular} {c|cc}
\hline\hline
\multirow{2}{*}{Parameters} & Mimick parameters from HEV sequence pair\\
& B1(Bur-82) vs B2(Bur-86)\\
\hline
\footnotesize Sequence length ($l$) & 8000\\
\footnotesize Probability of SNP ($p_{SNP}$) & 0.067\\
\footnotesize Probability of indel ($p_{indel}$) & 0.00014\\
\footnotesize Maximum length of indel ($I_{max}$) & 10\\
\footnotesize Zipfian distribution parameter ($s$) & 1.6\\
\hline\hline
\end{tabular}
\end{center}
\label{TAB:S2}
\end{table*}

\begin{table*}
\caption[Detailed simulation environment]{Detailed simulation environment}
\begin{center}
\begin{tabular} {c|ccccc}
\hline\hline
\multirow{2}{*}{Parameters} & Case 1 & Case 2-1 & Case 2-2 & Case 3 & Case 4\\
& Only SNP & Only indel ($I_{max}$) & Only indel ($p_{indel}$) & SNP\&indel & independent \\
\hline
\footnotesize Sequence length ($l$) & 8000 & 8000 & 8000 & 8000 & 8000 \\
\footnotesize Probability of SNP ($p_{SNP}$) & 0.05,0.1,0.2 & 0 & 0 & 0.1 & 1 \\
\footnotesize Probability of indel ($p_{indel}$) & 0 & 0.1 & 0.05,0.1,0.2 & 0.1 & 0  \\
\footnotesize Maximum length of indel ($I_{max}$) & 0 & 1,2,3 & 2 & 1,2,3 & 0  \\
\footnotesize Zipfian distribution parameter ($s$) & 1.6 & 1.6 & 1.6 & 1.6 & 1.6 \\
\hline\hline
\end{tabular}
\end{center}
\label{TAB:S3}
\end{table*}

\clearpage
\begin{table*}
\caption[BenchmarkedHEV genome sequences]{Benchmarked HEV genome sequences}
\begin{center}
\begin{tabular}{ccccc}
\hline\hline
NO. & STRAIN NAME & ACCESSION NO. & GENOTYPE & LENGTH\\
\hline
1 & B1 & M73218 & I & 7207\\
2 & B2 & D10330 & I & 7194\\
3 & I3 & AF076239 & I & 7194\\
4 & NP1 & AF051830 & I & 7199\\
5 & P2 & AF185822 & I & 7143\\
6 & Yam-67 & AF459438 & I & 7206\\
7 & C1 & D11092 & I & 7207\\
8 & C2 & L25595 & I & 7221\\
9 & C3 & L08816 & I & 7176\\
10 & C4 & D11093 & I & 7194\\
11 & China Hebei & M94177 & I & 7200\\
12 & P1 & M80581 & I & 7138\\
13 & I1 & X98292 & I & 7202\\
14 & Morocco & AY230202 & I & 7212\\
15 & T3 & AY204877 & I & 7170\\
16 & M1 & M74506 & II & 7180\\
17 & HE-JA10 & AB089824 & III & 7262\\
18 & JKN-Sap & AB074918 & III & 7256\\
19 & JMY-HAW & AB074920 & III & 7240\\
20 & SW-US1 & AF082843 & III & 7207\\
21 & US1 & AF060668 & III & 7202\\
22 & US2 & AF060669 & III & 7277\\
23 & JBOAR1-HYO04 & AB189070 & III & 7247\\
24 & JDEER-HYO03L & AB189071 & III & 7230\\
25 & JJT-KAN & AB091394 & III & 7218\\
26 & JIMO-HYO03L & AB189072 & III & 7180\\
27 & JRA1 & AP003430 & III & 7230\\
28 & JSO-HYO03L & AB189073 & III & 7180\\
29 & JTH-HYO03L & AB189074 & III & 7180\\
30 & JYO-HYO03L & AB189075 & III & 7180\\
31 & SWJ570 & AB073912 & III & 7257\\
32 & KYRGYZ & AF455784 & III & 7239\\
33 & ARKELL & AY115488 & III & 7255\\
34 & HE-JA1 & AB097812 & IV & 7258\\
35 & HE-JK4 & AB099347 & IV & 7250\\
36 & HE-JI4 & AB080575 & IV & 7186\\
37 & JAK-Sai & AB074915 & IV & 7236\\
38 & JKK-SAP & AB074917 & IV & 7235\\
39 & JSM-SAP94 & AB161717 & IV & 7202\\
40 & JSN-SAP-FH & AB091395 & IV & 7234\\
41 & JSN-SAP-FH02C & AB200239 & IV & 7251\\
42 & JTS-SAP02 & AB161718 & IV & 7202\\
43 & JYW-SAP02 & AB161719 & IV & 7202\\
44 & SWJ13-1 & AB097811 & IV & 7258\\
45 & SWCH25 & AY594199 & IV & 7270\\
46 & T1 & AJ272108 & IV & 7232\\
47 & CCC220 & AB108537 & IV & 7193\\
\hline\hline
\end{tabular}
\end{center}
\label{TAB:S4}
\end{table*}
\clearpage

\begin{table*}
\caption[Parameters used in the Clustal Omega software]{Parameters used in the Clustal Omega software}
\begin{center}
\begin{tabular} {c|c|c}
\hline\hline
Parameters & HEV simulation case & E.coli simulation case \\
\hline
K & \multirow{2}{*}{2} & \multirow{2}{*}{7} \\
\tiny (k-tuple) & &\\
\hline
signif & \multirow{2}{*}{4} & \multirow{2}{*}{500} \\
\tiny (Number of top diagonals to select) & &\\
\hline
window & \multirow{2}{*}{4} & \multirow{2}{*}{5} \\
\tiny (Allowable differences of diagonals nearby selected top diagonals) & &\\
\hline
wind\_gap & \multirow{2}{*}{5} & \multirow{2}{*}{5}  \\
\tiny (Allowable gaps for merging top diagonals) & &\\
\hline\hline
\end{tabular}
\end{center}
\label{TAB:S5}
\end{table*}

\begin{table*}
\caption[Parameters used in the MUMmer software]{Parameters used in the MUMmer software}
\begin{center}
\begin{tabular} {c|c|c}
\hline\hline
Parameters & HEV simulation case & E.coli simulation case \\
\hline
max\_gap & \multirow{2}{*}{90} & \multirow{2}{*}{100000} \\
\tiny (Maximum length of gaps which can be added into cluster) & &\\
\hline
min\_cluster & \multirow{2}{*}{20} & \multirow{2}{*}{10000} \\
\tiny (Minimum length of cluster) & &\\
\hline\hline
\end{tabular}
\end{center}
\label{TAB:S6}
\end{table*}

\end{document}